\documentstyle[12pt]{article}

\begin{document}
\baselineskip=18pt
\begin{titlepage}
\begin{center}
\large{\bf BILINEARIZATION OF SUPERSYMMETRIC KP HIERARCHIES ASSOCIATED WITH 
NON-TRIVIAL FLOWS} \end{center} \vspace{2 cm}
\begin{center}
{\bf Sasanka Ghosh}\footnote{email: sasanka@iitg.ernet.in} and
{\bf Debojit Sarma}\footnote{email: debojit@iitg.ernet.in}\\
{\it Department of Physics, Indian Institute of Technology,}\\
{\it North Guwahati, Guwahati 781039, INDIA}\\
\end{center}
\vspace{1 cm}


\vspace{3 cm}
\begin{abstract}
We show that the supersymmetric KdV and KP equations, related to the
non-trivial flows, can be cast in the Hirota bilinear form. The existence of
one, two and subsequently $N$-soliton solutions is explicitly demonstrated.
\end{abstract}
\end{titlepage}

\newpage

\section{Introduction}

The Hirota bilinear formalism is a powerful tool in computing multi-soliton solutions
and in establishing the integrability of nonlinear evolution equations. Integrable
hierarchies such as the KdV, the KP, and a host of other systems can be cast into the
bilinear form \cite{1,1a,1b,1c,1d,2,2a,2b,2c,2d,2e} and soliton solutions can be then
constructed by a simple algebraic procedure.

Supersymmetric generalizations of these systems have been formulated and particularly the
superKdV \cite{10,11} and the superKP \cite{3,9} have been extensively investigated. The integrability of the
superKP hierarchies in the Lax sense and on the basis their biHamiltonian structures is known.
The biHamiltonian structures of the superKdV as well as the superKP hierarchies have been
obtained by suitably formulating the Gelfand Dikii method for the supersymmetric systems.
In fact, most of the techniques used in the bosonic integrable systems have been extended to the
supersymmetric version of the theory. Mention may be made of the Backlund tranformations \cite{4},
the Painleve analysis \cite{5}, the Darboux transformations \cite{6} and the $\tau$-functions \cite{7}
vis a vis the soliton solutions which have been generalized to the supersymmetric framework. However,
only sporadic attempts have been made to extend the bilinear formalism to the supersymmetric
equations and those have been in $N=1$ supersymmetry. The $N=1$ supersymmetric version of
the KdV equation, as well as a few other supersymmetric equations have been written in bilinear form by means of 
supersymmetric analogue of the Hirota bilinear operators \cite{8}.

In this paper, we show that the supersymmetric KdV and KP equations associated with $N=2$ conformal
algebras can be bilinearized and the $N$ soliton solutions have been obtained for these systems.
This also establishes their integrability from an alternate point of view.

The equations of motion of the supersymmetric KP hierarchies are obtained from the
even parity superLax operator \cite{9}
\begin{equation}
L = D^{2} + \sum^{\infty }_{i=0} u_{i-1}(X) D^{-i}
\label{1.1}
\end{equation}
where, $D$ is the superderivative with $D^{2} = d/dx$ and $u_{i-1}(X)$ are superfields in
$X = (x,\theta )$ space, $\theta $ being Grassmann odd coordinate. The grading of $u_{i-1}(X)$
is $|u_{i-1}|= i$ so that $u_{2i-1}$ are bosonic superfields, whereas $u_{2i}$ are fermionic ones.
This choice of the superLax operator ensures that under the reduction
\begin{equation}
\tilde{L}_{n}=L^{n}_{>0}
\label{1.2}
\end{equation}
one is able to obtain the superLax operator used by Inami and Kanno  \cite{10} for
the generalized $N=2$ KdV from (\ref{1.1}). Use of the nonstandard flow equation
\begin{equation}
\frac{dL}{dt_{n}}=[L^{n}_{>0},L]
\label{1.3}
\end{equation}
(where, `$>0$' implies the +ve part of $L^{n}$ without $D^{0}$ term), in conjunction with
the Lax operator (\ref{1.1}), allows one to obtain the following evolution
equations, the first three being
\begin{eqnarray}
&&\frac{du_{i-1}}{dt_{1}}=u^{[2]}_{i-1}
\label{1.4}\\
&&\frac{du_{i-1}}{dt_{2}}=2u^{[2]}_{i+1}+u^{[4]}_{i-1}+2u_{0}u^{[1]}_{i-1}+2u_{-1}u^{[2]}_{i-1}
-2\left[\begin{array}{c}i+1\\1\end{array}\right]u_{i}u^{[1]}_{-1} \nonumber\\
&&-2(1+(-1)^{i})u_{0}u_{i}+2\sum^{i-1}_{m=0}\left[\begin{array}{c}i\\m+1\end{array}\right]
(-1)^{i+1+[-m/2]}u_{i-m-1}u^{[m+1]}_{0} \nonumber\\
&&+2\sum^{i-1}_{m=0}\left[\begin{array}{c}i+1\\m+2\end{array}\right](-1)^{[m/2]}u_{i-m-1}u^{[m+2]}_{-1}
\label{1.5}\\
&&\frac{du_{i-1}}{dt_{3}}=3u^{[2]}_{i+3}+3u^{[4]}_{i+1}+u^{[6]}_{i-1}
+6u_{-1}u^{[2]}_{i+1}+3u_{-1}u^{[4]}_{i-1} \nonumber\\
&&-3\left[\begin{array}{c}i+3\\1\end{array}\right]u_{i+2}u^{[1]}_{-1}+3
\left[\begin{array}{c}i+3\\2\end{array}\right]u_{i+1}u^{[2]}_{-1}
+3\left[\begin{array}{c}i+3\\3\end{array}\right]u_{i}u^{[3]}_{-1} \nonumber\\
&&-3(1+(-1)^{i})u_{0}u_{i+2}+3u_{0}u^{[1]}_{i+1}-3(-1)^{i}u_{0}u^{[2]}_{i}
+3u_{0}u^{[3]}_{i-1} \nonumber\\
&&+3\left[\begin{array}{c}i+2\\1\end{array}\right](-1)^{i}u_{i+1}u^{[1]}_{0}
-3\left[\begin{array}{c}i+2\\2\end{array}
\right](-1)^{i}u_{i}u^{[2]}_{0} \nonumber\\
&&+3(u_{1}+2u^{[2]}_{-1}+u^{2}_{-1})u^{[2]}_{i-1}
-3\left[\begin{array}{c}i+1\\1\end{array}\right]u_{i}(u_{1}+u^{[2]}_{-1}+u^{2}_{-1})^{[1]} \nonumber\\
&&+3(u_{2}+2u_{-1}u_{0}+u^{[2]}_{0})u^{[1]}_{i-1}
-3(1+(-1)^{i})(u_{2}+2u_{-1}u_{0}+u^{[2]}_{0})u_{i} \nonumber\\
&&-3\sum^{i-1}_{m=0}\left[\begin{array}{c}i+3\\m+4\end{array}\right](-1)^{[m/2]}u_{i-m-1}u^{[m+4]}_{-1} \nonumber\\
&&-3\sum^{i-1}_{m=0}\left[\begin{array}{c}i+2\\m+3\end{array}\right](-1)^{i+[-m/2]}u_{i-m-1}u^{[m+3]}_{0} \nonumber\\
&&+3\sum^{i-1}_{m=0}\left[\begin{array}{c}i+1\\m+2\end{array}\right](-1)^{[m/2]}u_{i-m-1}
(u_{1}+u^{[2]}_{-1}+u^{2}_{-1})^{[m+2]} \nonumber\\
&&+3\sum^{i-1}_{m=0}\left[\begin{array}{c}i\\m+1\end{array}\right](-1)^{i+[-m/2]}u_{i-m-1}
(u_{2}+2u_{-1}u_{0}+u^{[2]}_{0})^{[m+1]}
\label{1.6}
\end{eqnarray}
In the above equations $t_1$ may be identified with the $x$ coordinate, whereas $t_2$
may be associated with the $y$ coordinate. The $t_{3}$ evolution, on the other hand,
may be identified with the first time evolution for the superfields $u_{i-1}$ for the $N=2$
supersymmetric KP hierarchies. It is interesting to note that if the superfields are made
independent of the $y$ coordinate in (\ref{1.5},\ref{1.6}), the superKP equations reduce to
the supersymmetric KdV equations of Inami Kanno type \cite{11}. A major consequence of the
nonstandard flow equation is that the lowest field $u_{-1}$ possesses non-trivial dynamics.
This property of the lowest field, which is absent in the standard flow, is instrumental in the
associated Poisson bracket structure being local. Moreover, it is the $u_{-1}$ field which
provides the spin one current of the $N=2$ superconformal algebra.

The plan of the paper is as follows. In section II, we cast the $N=2$ KdV and KP equations in the
Hirota bilinear form. In the section III, the  existence of one and two soliton solutions are
demonstrated. The $N$ soliton solution for the equations is discussed in section IV. Section V
forms the conclusion.

\setcounter{equation}{0}

\section{The Hirota bilinear form}

The $N=2$ superKdV equations proposed by Inami Kanno \cite{11} may be obtained from (\ref{1.5})
and (\ref{1.6}) under the reduction that the fields are independent of the $y$ coordinate.
The equations of motion for the superfields $u_{-1}$ and $u_{0}$ eventually become
\begin{equation}
\partial_{t}u_{-1}+u^{[6]}_{-1}+3\left(u^{[1]}_{-1}u_{0}\right)^{[2]}
-\frac{1}{2}\left(u^{3}_{-1}\right)^{[2]}=0
\label{2.1}
\end{equation}
\begin{equation}
\partial_{t}u_{0}+u^{[6]}_{0}-3\left(u_{0}u^{[1]}_{0}\right)^{[2]}-
\frac{3}{2}\left(u_{0}u^{2}_{-1}\right)^{[2]}+3\left(u_{0}u^{[2]}_{-1}\right)^{[2]}=0
\label{2.2}
\end{equation}
where $[\;n\;]$ denotes the $n$th derivative with respect to the $D$ operator, defined by
\begin{equation}
D=\frac{\partial}{\partial\theta}+\theta\frac{\partial}{\partial x}
\label{2.3}
\end{equation}
In order to cast the equations (\ref{2.1},\ref{2.2}) in the bilinear form, we write the
superfields in the form
\begin{equation}
u_{-1}=2D^{2}\log\frac{\tau_{1}}{\tau_{2}}
\label{2.4}
\end{equation}
for $u_{-1}$ and
\begin{equation}
u_{0}=2D^{3}\log\frac{\tau_{1}}{\tau_{2}}
\label{2.5}
\end{equation}
for $u_{0}$, where $\tau_{1}$ and $\tau_{2}$ are bosonic superfields. Interestingly, under the transformations
(\ref{2.4},\ref{2.5}), both the equations (\ref{2.1}) and (\ref{2.2}) reduce to a single equation
in terms of the $\tau$ functions as
\begin{equation}
\partial_{t}\log\frac{\tau_{1}}{\tau_{2}}+\partial^{3}_{x}\log\frac{\tau_{1}}{\tau_{2}}
-2(\partial_{x}\log\frac{\tau_{1}}{\tau_{2}})^{3}=0
\label{2.6}
\end{equation}
Notice that in terms of $\tau$ functions, the dynamical equations do not possess the
supersymmetric derivative operator, $D$. This reflects the fact that the $\tau$ functions
are essentially bosonic in nature. In terms of the Hirota derivative, defined by
\begin{equation}
{\bf{D}}^{n}_{x}f.g=(\partial_{x_{1}}-\partial_{x_{2}})^{n}f(x_{1})g(x_{2})|_{x_{1}=x_{2}=x}
\label{2.9}
\end{equation}
the equation (\ref{2.6}) yields the bilinear forms as
\begin{equation}
({\bf{D}}_{t}+{\bf{D}}^{3}_{x})(\tau_{1}.\tau_{2})=0
\label{2.7}
\end{equation}
\begin{equation}
{\bf{D}}^{2}_{x}(\tau_{1}.\tau_{2})=0
\label{2.8}
\end{equation}
We will see that the $N=2$ superKP equations also contain bosonic derivatives 
only when expressed in terms of the $\tau$ functions.

The dynamical equations for the superfields $u_{-1}$ and $u_{0}$ constitute the $N=2$ KP equations.
It follows from (\ref{1.5}) and (\ref{1.6}) that the dynamical equations of the superfields
$u_{-1}$ and $u_{0}$ become nonlocal. The explicit forms of the equations may be given by
\begin{eqnarray}
&&\partial_{t}u_{-1}-\frac{1}{4}u^{[6]}_{-1}+\frac{1}{2}\left(u^{3}_{-1}\right)^{[2]}
-\frac{3}{2}\left(u_{0}u^{[1]}_{-1}\right)^{[2]}-\frac{3}{4}\partial^{2}_{y}u^{[-2]}_{-1} \nonumber\\
&&-\frac{3}{2}u^{[2]}_{-1}\partial_{y}u^{[-2]}_{-1}
+\frac{3}{2}u^{[1]}_{-1}\partial_{y}u^{[-2]}_{0}
+\frac{3}{2}u_{0}\partial_{y}u^{[-1]}_{-1}-3u_{0}\partial_{y}u^{[-2]}_{0}=0
\label{2.10}
\end{eqnarray}
\begin{eqnarray}
&&\partial_{t}u_{0}-\frac{1}{4}u^{[6]}_{0}-\frac{3}{2}\left(u_{0}u^{[1]}_{0}\right)^{[2]}
+\frac{3}{2}\left(u_{0}u^{[2]}_{-1}\right)^{[2]}+\frac{3}{2}\left(u_{0}u^{2}_{-1}\right)^{[2]} \nonumber\\
&&-\frac{3}{4}\partial^{2}_{y}u^{[-2]}_{0}
-\frac{3}{2}\left(u_{0}\partial_{y}u^{[-2]}_{0}\right)^{[1]}
-\frac{3}{2}u^{[2]}_{0}\partial_{y}u^{[-2]}_{-1}
-\frac{3}{2}u_{0}\partial_{y}u_{-1}=0
\label{2.11}
\end{eqnarray}
If we now rewrite the equations (\ref{2.10},\ref{2.11}) in terms of the $\tau$ functions using the definitions
\begin{equation}
u_{-1}=D^{2}\log\frac{\tau_{1}}{\tau_{2}}
\label{2.11a}
\end{equation}
\begin{equation}
u_{0}=D^{3}\log\frac{\tau_{1}}{\tau_{2}}
\label{2.11b}
\end{equation}
both the superKP equations are transformed to a single equation
as before and become bosonic in nature. In terms of the $\tau$ functions, (\ref{2.10}) and (\ref{2.11})
reduce to
\begin{eqnarray}
&&\partial_{t}\partial_{x}\log\frac{\tau_{1}}{\tau_{2}}-\frac{1}{4}\partial_{x}^{4}\log\frac{\tau_{1}}{\tau_{2}}
+\frac{1}{2}\partial_{x}\left(\partial_{x}\log\frac{\tau_{1}}{\tau_{2}}\right)^{3} \nonumber\\
&&-\frac{3}{4}\partial^{2}_{y}\log\frac{\tau_{1}}{\tau_{2}}-\frac{3}{2}\left(\partial^{2}_{x}
\log\frac{\tau_{1}}{\tau_{2}}\right)\left(\partial_{y}\log\frac{\tau_{1}}{\tau_{2}}\right)=0
\label{2.14}
\end{eqnarray}
Rescaling (\ref{2.14}) by $t=4t$, $x=-x$, and $y=\frac{1}{2}y$, we have
\begin{eqnarray}
&&\partial_{t}\partial_{x}\log\frac{\tau_{1}}{\tau_{2}}+\partial_{x}^{4}\log\frac{\tau_{1}}{\tau_{2}}
-2\partial_{x}\left(\partial_{x}\log\frac{\tau_{1}}{\tau_{2}}\right)^{3} \nonumber\\
&&+12\partial^{2}_{y}\log\frac{\tau_{1}}{\tau_{2}}+12\left(\partial^{2}_{x}
\log\frac{\tau_{1}}{\tau_{2}}\right)\left(\partial_{y}\log\frac{\tau_{1}}{\tau_{2}}\right)=0
\label{2.15}
\end{eqnarray}
The rescaling of (\ref{2.14}) and the change of the multiplying factor from two to unity in (\ref{2.11a})
and (\ref{2.11b}) is done so that the bilinear forms reduces to the known ones
in the bosonic limit \cite{2,2a}. Finally (\ref{2.15}) can be cast into the Hirota bilinear forms as
\begin{equation}
({\bf{D}}_{x}{\bf{D}}_{t}+{\bf{D}}^{4}_{x}+12{\bf{D}}^{2}_{y})(\tau_{1}.\tau_{1})=0
\label{2.16}
\end{equation}
\begin{equation}
({\bf{D}}_{x}{\bf{D}}_{t}+{\bf{D}}^{4}_{x}+12{\bf{D}}^{2}_{y})(\tau_{2}.\tau_{2})=0
\label{2.17}
\end{equation}
\begin{equation}
({\bf{D}}^{2}_{x}-2{\bf{D}}_{y})(\tau_{1}.\tau_{2})=0
\label{2.18}
\end{equation}

\setcounter{equation}{0}

\section{One and two soliton solutions}

In this section we demonstrate the existence of one and two soliton solutions for superKP as
well as superKdV equations.

(i)\underline{The superKdV equation}

The general form of the tau-functions for the one soliton solution of the superKdV equation may be written as
\begin{equation}
\tau_{1}=1+\alpha e^{\eta}
\label{3.1}
\end{equation}
\begin{equation}
\tau_{2}=1+\beta e^{\eta}
\label{3.2}
\end{equation}
$\alpha$ and $\beta$ in (\ref{3.1},\ref {3.2}) are nonzero constants and
\begin{equation}
\eta=kx+\omega t+\zeta \theta
\label{3.3}
\end{equation}
where $k$ and $\omega$ are the bosonic parameters and $\zeta$ is the Grassmann odd parameter. 
Substituting (\ref{3.1}) and (\ref{3.2}) back into the Hirota form of the superKdV equation (\ref{2.8}),
we find the non-trivial solutions provided
\begin{equation}
\beta=-\alpha.
\label{3.4a}
\end{equation}
The dispersion relation, however, follows from (\ref{2.7}) as
\begin{equation}
\omega+k^{3}=0
\label{3.4}
\end{equation}
which is identical to the bosonic KdV or mKdV equations. The fermionic 
parameter, $\zeta$, however, remains arbitrary at the one soliton level.

The two soliton solution of the superKdV equations may be obtained by writing the tau functions in the form
\begin{equation}
\tau_{1}=1+\alpha_{1}e^{\eta_{1}}+\alpha_{2}e^{\eta_{2}}+\alpha_{1}\alpha_{2} A_{12}e^{\eta_{1}+\eta_{2}}
\label{3.5}
\end{equation}
and
\begin{equation}
\tau_{2}=1+\beta_{1}e^{\eta_{1}}+\beta_{2}e^{\eta_{2}}+\beta_{1}\beta_{2} B_{12}e^{\eta_{1}+\eta_{2}}
\label{3.6}
\end{equation}
where
\begin{equation}
\eta_{1}=k_{1}x+\omega_{1} t+\zeta_{1}\theta
\label{3.7}
\end{equation}
\begin{equation}
\eta_{2}=k_{2}x+\omega_{2} t+\zeta_{2}\theta
\label{3.8}
\end{equation}
In (\ref{3.7},\ref{3.8}) the parameters $k_1$, $k_2$, $\omega_1$ and $\omega_2$ are bosonic, while
$\zeta_1$ and $\zeta_2$ are fermionic ones, as before.

Substituting the above equations (\ref{3.5}-\ref{3.8}), in the second bilinear form (\ref{2.8})
leads to a set of conditions among the parameters. Like the one soliton case, the nontrivial
solutions of the bilinear form yields $\beta_{i}=-\alpha_{i}$, where $i=1,2$.
The coefficient of the bilinear term $e^{\eta_{1}+\eta_{2}}$, on the other hand, determines $A_{12}$
in terms of $k_i$ as
\begin{equation}
A_{12}=B_{12}=\frac{(k_{1}-k_{2})^2}{(k_{1}+k_{2})^2}
\label{3.9}
\end{equation}
Finally, the coefficient of the trilinear terms like $e^{2\eta_{1}+\eta_{2}}$ or $e^{\eta_{1}+2\eta_{2}}$
gives rise to $A_{12}=B_{12}$.
As earlier, the dispersion relations are derivable from (\ref{2.7}). This leads to the two independent
dispersion relations, one involving the parameters of $\tau_1$ and the other being the parameters of
$\tau_2$. The explicit forms of the dispersion relations
are given by
\begin{equation}
\omega_{1}+k^{3}_{1}=0
\label{3.10}
\end{equation}
and
\begin{equation}
\omega_{2}+k^{3}_{2}=0
\label{3.11}
\end{equation}
The parameters $\alpha_i$ are abitrary even for the two soliton solutions. we will see that
the parameters $\alpha_i$ remain undetermined even for the $N$ soliton solutions. At the two
soliton level also the fermionic parameters $\zeta_1$ and $\zeta_2$ remain arbitrary.

(ii)\underline{The superKP equation}

To find the one soliton solution of the superKP equation, we use the the same form
of the $\tau$ functions, namely,
\begin{equation}
\tau_{1}=1+\alpha e^{\eta}
\label{3.12}
\end{equation}
\begin{equation}
\tau_{2}=1+\beta e^{\eta}
\label{3.13}
\end{equation}
but the parameter $\eta$ now depends also on the $y$ coordinate as follows.
\begin{equation}
\eta=k_{x}x+k_{y}y+\omega t +\zeta \theta
\label{3.14}
\end{equation}
It is easily shown that plugging these back in first two equations, namely (\ref{2.16}) and (\ref{2.17})
of the Hirota form of the KP, the dispersion relation may be obtained as
\begin{equation}
\omega k_{x}+k^{4}_{x}+12k^{2}_{y}=0
\label{3.15}
\end{equation}
From the third equation of this set (\ref{2.18}), $k_{y}$ is related to $k_{x}$ through
\begin{equation}
k_{y}=\frac{1}{2}\frac{(\alpha+\beta)}{(\alpha-\beta)}k^{2}_{x}
\label{3.16}
\end{equation}
This imposes the following constraint between $\alpha$ and $\beta$ parameters like
\begin{equation}
\beta\neq\alpha
\label{3.17}
\end{equation}
Notice that for the superKP equation the condition on $\alpha$ and $\beta$ parameters are less restrictive.
However, if the further restriction $\beta=-\alpha$ is imposed on the parameters, evidently $k_y$
in (\ref{3.17}) becomes zero and as a consequence the one soliton solution reduces to the superKdV one (\ref{3.4}).
This reduction, in fact, will be observed for all the $N$ soliton solutions.

For the two soliton solution for the superKP equation, the tau functions may be written once again as
(\ref{3.5}) and (\ref{3.6}), with $\eta_{1}$ and $\eta_{2}$ being now dependent on the $y$ coordinate.
The explicit forms of $\eta_{1}$ and $\eta_{2}$ are given by
\begin{equation}
\eta_{1}=k_{1x}x+k_{1y}y+\omega_{1}t +\zeta_1 \theta
\label{3.18}
\end{equation}
\begin{equation}
\eta_{2}=k_{2x}x+k_{2y}y+\omega_{2}t + \zeta_2 \theta
\label{3.19}
\end{equation}
We will obtain two dispersion relations corresponding to the the parameters $\eta_1$ and $\eta_2$.
The dispersion relations resulting from (\ref{2.16}) and (\ref{2.17}) become
\begin{equation}
\omega k_{ix}+k^{4}_{ix}+12k^{2}_{iy}=0,\;\;\;\;\;i=1,2
\label{3.20}
\end{equation}
The expressions for $A_{12}$ and $B_{12}$ may be obtained from (\ref{2.18}) by a procedure
identical to the one used for the superKdV equation. It turns out that
\begin{equation}
A_{12}=B_{12}
\label{3.22}
\end{equation}
and $A_{12}$ yields
\begin{equation}
A_{12}=-\frac{(\alpha_{1}\beta_{2}+\alpha_{2}\beta_{1})(k_{1x}-k_{2x})^2
-2(\alpha_{1}\beta_{2}-\alpha_{2}\beta_{1})(k_{1y}-k_{2y})}
{(\alpha_{1}\alpha_{2}+\beta_{1}\beta_{2})(k_{1x}+k_{2x})^2
-2(\alpha_{1}\alpha_{2}-\beta_{1}\beta_{2})(k_{1y}+k_{2y})}
\label{3.21}
\end{equation}
Quite clearly, (\ref{3.21}) reduces to the analogous expression for KdV in the limit
$\beta_{i}=-\alpha_{i}$ $(i=1,2)$ and the constraint (\ref{3.17}) holds good for two solitons also.

\setcounter{equation}{0}

\section{N soliton solution}

It is well known that the existence of one and two soliton solutions do not ensure the exact
integrability by Hirota method. In fact, the method of obtaining one and two soliton solutions
considered in the previous section always leads to the nontrivial solutions if the equations of
motion can be cast in the bilinear forms. This method rather determines the unknown parameters
involved in the trial solutions upto two soliton solutions. It is thus necessary to show the
existence of atleast three soliton solution, which checks the consistency of the solution.
In this section we discuss the $N$ soliton solutions and obtain the three soliton solutions in particular.
The $\tau$ function for the $N$ soliton of both superKdV and superKP may be written as
\begin{equation}
\tau_{1}=\sum_{\mu_{i}=0,1}exp\left(\sum_{i,j=1}^{N}\phi (i,j)\mu_{i}\mu_{j}+\sum_{i=1}^{N}\mu_{i}(\eta_{i}+\log\alpha_{i})\right)
\;\;\;\;\;(i<j)
\label{4.1}
\end{equation}
and $A_{ij}=e^{\phi (i,j)}$. For $\tau_{2}$ we replace $\alpha_{i}$ by $\beta_{i}$ and $A_{ij}$ by $B_{ij}$.
For example, the explicit forms of $\tau_1$ and $\tau_2$ for the three soliton solution become
\begin{eqnarray}
&&\tau_{1}=1+\alpha_{1}e^{\eta_{1}}+\alpha_{2}e^{\eta_{2}}+\alpha_{3}e^{\eta_{3}}
+\alpha_{1}\alpha_{2} A_{12}e^{\eta_{1}+\eta_{2}}+\alpha_{1}\alpha_{3} A_{13}e^{\eta_{1}+\eta_{3}}\nonumber\\
&&+\alpha_{2}\alpha_{3} A_{23}e^{\eta_{2}+\eta_{3}}+\alpha_{1}\alpha_{2}\alpha_{3} A_{12}A_{13}A_{23}
e^{\eta_{1}+\eta_{2}+\eta_{3}}
\label{4.2}
\end{eqnarray}
and
\begin{eqnarray}
&&\tau_{2}=1+\beta_{1}e^{\eta_{1}}+\beta_{2}e^{\eta_{2}}+\beta_{3}e^{\eta_{3}}
+\beta_{1}\beta_{2} B_{12}e^{\eta_{1}+\eta_{2}}+\beta_{1}\beta_{3} B_{13}e^{\eta_{1}+\eta_{3}}\nonumber\\
&&+\beta_{2}\beta_{3} B_{23}e^{\eta_{2}+\eta_{3}}+\beta_{1}\beta_{2}\beta_{3} B_{12}B_{13}B_{23}
e^{\eta_{1}+\eta_{2}+\eta_{3}}
\label{4.3}
\end{eqnarray}
Notice that the three soliton solution does not contain any new unknown parameter; it is expressed
in terms of the parameters of the two soliton solutions only. These forms of three soliton solution
give rise to a set of conditions among the parameters, which determine the consistency of the solutions.
For superKP equation, $\eta_i$ for $i=1,2,3$, depends on the $y$ coordinate also, the explicit form being
\begin{equation}
\eta_i= k_{ix}x+k_{iy}y+\omega_i t+\zeta_i \theta
\label{4.4}
\end{equation}
If we substitute the form of $A_{ij}$ for $i<j$ from (\ref{3.21}) as
\begin{equation}
A_{ij}=B_{ij}= -\frac{(\alpha_{i}\beta_{j}+\alpha_{j}\beta_{i})(k_{ix}-k_{jx})^2
-2(\alpha_{i}\beta_{j}-\alpha_{j}\beta_{i})(k_{iy}-k_{jy})}
{(\alpha_{i}\alpha_{j}+\beta_{i}\beta_{j})(k_{ix}+k_{jx})^2
-2(\alpha_{i}\alpha_{j}-\beta_{i}\beta_{j})(k_{iy}+k_{jy})}
\label{4.5}
\end{equation}
and the dispersion relation corresponding to each $\eta_i$ from (\ref{3.19}) as
\begin{equation}
\omega k_{ix}+k^{4}_{ix}+12k^{2}_{iy}=0
\label{4.6}
\end{equation}
in the equations (\ref{4.2}) and (\ref{4.3}), $\tau_1$ and $\tau_2$ eventually satisfy the
equations (\ref{2.16},\ref{2.17},\ref{2.18}), provided $k_{ix}$ and $k_{iy}$ are related
as earlier like
\begin{equation}
k_{iy}=\frac{1}{2}\frac{(\alpha_i+\beta_i)}{(\alpha_i-\beta_i)}k^{2}_{ix}
\label{4.7}
\end{equation}
This, indeed, ensures the existence of three soliton solution for the superKP equation.
It follows from (\ref{4.7}) straightforwardly that $\beta_{i}\neq\alpha_{i}$ for a consistent
solution. The three soliton solution for the superKdV equation may be obtained from
(\ref{4.5}-\ref{4.7}) by imposing the conditions $\alpha_{i}=-\beta_{i}$, which make the solutions
independent of the $y$ coordinate.

From the form of the trial solution (\ref{4.1}) for the $N$ soliton solution, it is evident that the $N$
soliton solution may be written in terms of $\alpha_i$ and $\beta_i$ as well as $A_{ij}$ and $B_{ij}$,
whose explicit forms may be obtained from (\ref{4.5}). All the higher soliton solutions starting from
the three soliton solution thus lead to a set of constraint conditions on the polynomials of the
parameters $k_{ix}$, $k_{iy}$ and $\omega_i$. This puts a stringent condition on the solution space.
To be explicit, it is found that for both the superKdV and superKP equations that the terms of different
degrees in $\eta_1$ and $\eta_2$ give rise to different conditions, where by degree of a term we define
the number of times $\eta_1$ and $\eta_2$ are present in a term. For example, the terms of degree one and
degree $(2N-1)$ in (\ref{2.18}) put the condition on $k_{ix}$ and $k_{iy}$ as in (\ref{4.7}).
The terms of the same degrees in (\ref{2.16}) and (\ref{2.17}), on the other hand, determine the
dispersion relations as in (\ref{4.6}). $A_{ij}$ are evaluated in terms of $k_{ix}$ and $k_{iy}$
from terms of degree two and $(2N-2)$ like in (\ref{4.5}), whereas the conditions $A_{ij}=B_{ij}$ for $i<j$ follow from
degree three and degree $(2N-3)$ terms. The terms of other degrees
give rise to a set of polynomials in $\omega_i$, $k_{ix}$ and $k_{iy}$, which become zero identically.
In general, if $N$ is odd, the coefficients of even degree terms become trivial. The first non-trivial
term of odd degree appears at the three soliton level, which is of the form $e^{\eta_1+\eta_2+\eta_3}$.
If $N$ is even, the terms of odd degree, however, become trivial.

\setcounter{equation}{0}

\section{Conclusion}

It was shown that the $N=2$ supersymmetric KdV and KP equations associated with nontrivial flows are
bilinearizable. The bilinear forms can be written in terms of the Hirota derivative, but the the $\tau$
functions become bosonic superfields. Both these supersymmetric equations are shown to possess one,
two and three soliton solutions. The existence of three soliton solutions ensure the exact integrability
of both the dynamical equations. The $N$ soliton solutions are also discussed for both
systems. It is found that under the restrictions $\alpha_i=-\beta_i$, the superKP equation reduces
to the superKdV equation. It is interesting to note that in the bosonic limit, the superKdV and the
superKP equations  become isomorphic to the bosonic modified KdV and the modified KP equations respectively.
Since the fermionic parameters do not play a significant role in this case, the solution space of the superKdV
as well as the superKP equation becomes bosonic in nature. On the other hand, 
the solution space of the superKP hierarchies
associated with the standard flows is quite different from those of the non-standard flows. This result
will be published elsewhere.

\vspace{1 cm}

{\it SG would like to thank DST, Govt. of India for financial support under the project no. 100/(IFD)/2066/2000-2002.}

\end{document}